# Stable Self-Charged Perovskite Quantum Rods for Liquid Laser with Near-Zero Threshold

*Jialu Li[1,†], Xue Han[1,†], Wenjie Wang[2], Jinhui Wang[1], Tingting Zhang[2], Yuting Wu[3], Guofeng Zhang[1\*], Bin Li[1], Changgang Yang[1], Wenli Guo[1], Mi Zhang[1], Ruiyun Chen[1], Chengbing Qin[1], Jianyong Hu[1], Zhichun Yang[1], Shaoding Liu[2\*], Yue Wang[3\*], Yunan Gao[4\*], Jie Ma[1], Liantuan Xiao[1\*], Suotang Jia[1]*

1. State Key Laboratory of Quantum Optics Technologies and Devices, Institute of Laser Spectroscopy, Collaborative Innovation Center of Extreme Optics, Shanxi University, Taiyuan, 030006, China
2. Key Lab of Advanced Transducers and Intelligent Control System of Ministry of Education, Taiyuan University of Technology, 79 Yingze Street, Taiyuan, 030024, China
3. School of Microelectronics, College of Materials Science and Engineering, Nanjing University of Science and Technology, Nanjing 210094, China
4. State Key Laboratory for Mesoscopic Physics and Frontiers Science Center for Nano-optoelectronics, School of Physics, Peking University, Beijing 100871, China

[†] These authors contributed equally to this work.

\*Corresponding Authors.

Email: guofeng.zhang@sxu.edu.cn

liushaoding@tyut.edu.cn

ywang@njust.edu.cn

gyn@pku.edu.cn

xlt@sxu.edu.cn

**Abstract**

Colloidal quantum dots (QDs) are promising optical gain materials that require further threshold reduction to realize their full potential. While QD charging theoretically reduces the threshold to zero, its effectiveness has been limited by strong Auger recombination and unstable charging. Here we theoretically reveal the optimal combination of charging number and Auger recombination to minimize the lasing threshold. Experimentally, we develop stable self-charged perovskite quantum rods (QRs) as an alternative to QDs via state engineering and Mn-doping strategy. An unprecedented two-order-of-magnitude reduction in nonradiative Auger recombination enables QRs to support a sufficient charging number of up to 6. The QR liquid lasing is then achieved with a near-zero threshold of 0.098 using quasi-continuous pumping of nanosecond pulses, which is the lowest threshold among all reported QD lasers. These achievements demonstrate the potential of the specially engineered QRs as an excellent gain media and pave the way for their prospective applications.

Keywords: liquid laser, perovskite quantum rods, stable charging, nonradiative Auger recombination, near-zero lasing threshold

Colloidal semiconductor quantum dots (QDs) have demonstrated significant promise as laser gain media, with a plethora of research efforts focused on harnessing their unique optoelectronic characteristics.[1–4] These include a low lasing threshold, high thermal stability, resistance to photobleaching, and tunability across the entire visible spectrum.[1,2] The great promise of QDs has been recently reaffirmed by the achievement of the electrical-pumped amplified spontaneous emission.[5] However, their prospective applications require further achievement in preventing overheating and lowering the lasing thresholds. Recently, the successful demonstration of the high performance of liquid lasers based on rapidly heat dissipating QD solutions offers a promising platform.[6,7] Additionally, the solution-based gain media enables more convenient tuning of laser characteristics such as wavelength and spatial mode. These advantages make QD liquid lasers attractive for applications in astronomy, medicine and spectroscopy, including but not limited to optofluidic-based on-chip environmental monitoring, medical diagnostics, and chemical weapon detection.[8,9]

The early efforts to develop QD liquid lasers were unsuccessful due to the unexpectedly strong Auger loss of QDs.[10] The nonradiative transfer of exciton energy to a third carrier instead of photon emission,[1] the Auger process, inevitably increases the critical volume fraction required for achieving the laser regime to 0.2% (not readily accessible in QD solutions).[10] As a result, over the last two decades, numerous lasing demonstrations of QDs are limited to densely packed QD films.[3,4,10] Although some studies have achieved lasing regime in QD solutions by introducing high-quality optical resonant cavities, their lasing thresholds remain significantly higher than those of QD film lasers.[6,7,11–17] To overcome these challenges and unlock the full potential of QDs as gain media, further research is needed to develop strategies for achieving lower lasing thresholds.

For threshold reduction, the suppression of nonradiative Auger recombination is imperative.[1,3,4] The development of Auger-engineered QDs has led to an biexciton Auger lifetime of 2 ns,[3–5] lowering the threshold in QD film lasers. Nevertheless, their Auger process is still more efficient than their radiative process and requires further reduction. In addition to reducing Auger recombination, the lasing threshold can be further lowered by QD charging, which suppress ground-state absorption by converting biexciton gain into charged exciton gain.[1,3,4] Although optical gain and lasing have been achieved in charged QDs by photochemical[3,4] or electrochemical[18] charging approaches, these approaches involving chemical treatments can cause permanent damage to the QD gain material, leading to significant deterioration of its lasing efficiency.[3,18] Moreover, these approaches require continuous operation to maintain charging state, as spontaneous discharging processes occur naturally to lose charges.[3,4] Furthermore, QD charging dramatically increases Auger recombination by introducing additional Auger nonradiative pathways.[19] This limits the charging number of QD to usually 2 in previous works,[3,18] as too many charges would completely quench the photoluminescence (PL) and thereby the lasing of QDs, contradicting the expected positive effects to QD lasing.

In this study, we theoretically examine the optimal combination of charging number and Auger recombination to minimize the lasing threshold. Based on this insight, we design and realize one-dimensional $CsPbBr_3$ perovskite quantum rods (QRs) with Auger recombination reduced to an unprecedented two order of magnitude. Bromine interstitial ($Br_i^-$) states are ingeniously created in QRs to achieve a non-destructive self-charging effect of QRs. To stabilize QR charging, we develop Mn-doping strategy to preserve the $Br_i^-$ states by maximizing the energy barrier for $Br_i^-$ migration via hybridization of Mn 3d orbitals with Pb 6s-Br 4p antibonding states. The Mn-doping strategy enables a long-term QR self-charging with tunable charging number. Utilizing Mn-doped QRs charged with 6 electrons, a record low threshold ($\langle N \rangle$ ~ 0.098) in liquid lasing is reached using quasi-continuous pumping of nanosecond pulses.

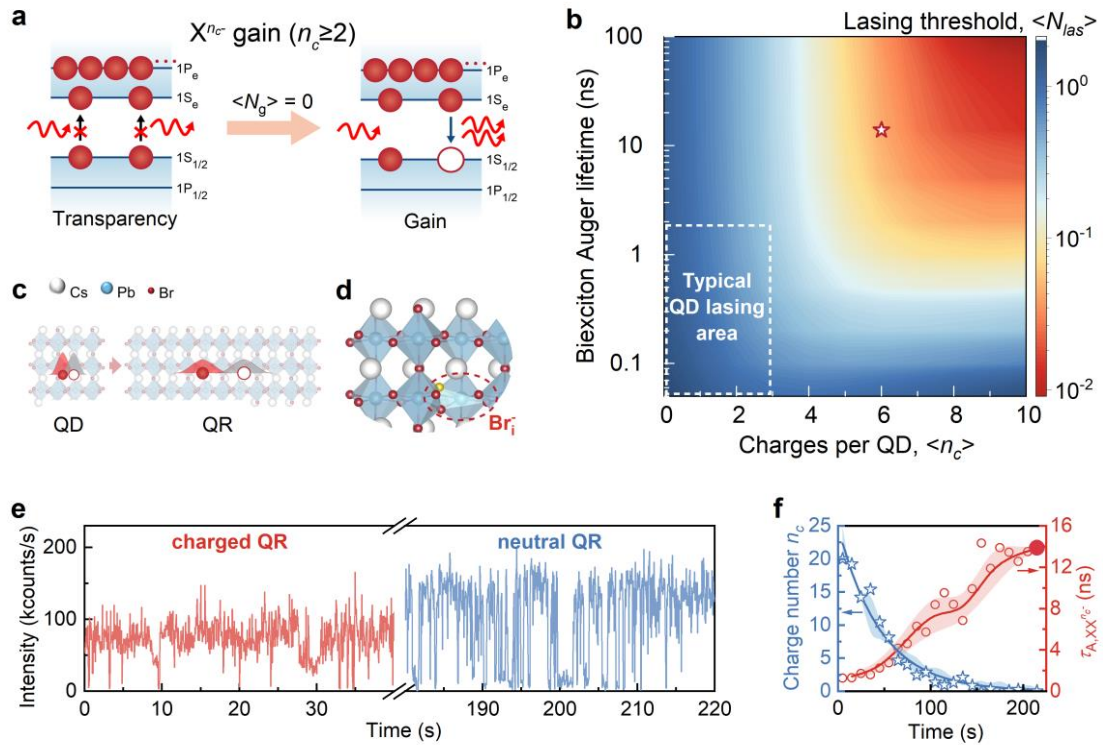

**Fig. 1 | Design and characterizations of CsPbBr₃ perovskite quantum rods (QRs) with self-charging effect and reduced Auger recombination.** (a) Schematics of optical-gain mechanisms and the optical-gain thresholds $\langle N_g \rangle$ for multi-charged exciton ($X^{n_c-}$) gain in multi-charged quantum dots (QDs) ($n_c \geq 2$). (b) Dependence of the lasing threshold $\langle N_{las} \rangle$ on the charging number per QD $\langle n_c \rangle$ and the Auger recombination (characterized by biexciton Auger lifetime $\tau_{A,XX}$ in neutral QDs). Limited by the charging number and the

Auger recombination of the typical QDs, their lasing thresholds are located within the white dashed frame. The red star represents the lasing threshold of the QR material with charging number of 6 and Auger lifetime of 13.9 ns in the work. (c) Top view of CsPbBr$_3$ perovskite QDs (left) and QRs (right). The one-dimensional elongated structure of QR reduces the Auger recombination. (d) Schematic depiction of the bromine interstitial state ($Br_i^-$) as a hole acceptor for the self-charging effect in the QR by creating a CsBr-rich environment. Excess Cs$^+$, which are automatically added to the surface of the QR in its natural position ($Cs_{Cs}^+$) to balance the charges of $Br_i^-$, are not considered as states and are not shown for simplicity. (e) Typical PL intensity time trajectory for a single CsPbBr$_3$ QR at excitation condition $\langle N \rangle = 0.2$. $\langle N \rangle$ represents average exciton number per QD. The red and blue regions represent the time frames from 0 to 40 s and from 180 to 220 s of the PL trajectory, respectively, and the whole PL trajectory from 0 to 220 s is presented in Supplementary Fig. 7. (f) Evolutions of charging number $n_c$ (blue stars) in the self-charged QR and corresponding biexciton Auger lifetime (red circle) over time. Errors are indicated by shaded areas. The constant decay of $n_c$ is caused by the loss of hole acceptors.

**Synergistic effects of Auger and QD charging on lasing threshold**

Since the band-edge states of QDs are twofold degenerate[20], optical gain in QDs is achieved by biexciton gain, with an optical-gain threshold $\langle N_g \rangle$ value of 1 in the simplified model (Supplementary Fig. 1a).[3] By charging the QD with an extra electron ($n_c = 1$), the ground-state absorption is partially blocked, and $\langle N_g \rangle$ decreases sharply to 0.5 (Supplementary Fig. 1b).[3,4] When the charging number $n_c \geq 2$, the ground-state absorption is completely blocked, and $\langle N_g \rangle$ decreases sharply to 0 (Fig. 1a), achieved by the multi-charged exciton ($X^{n_c-}$) gain.[3,4] However, in real situations, both the exciton number $N$ and the charge number $n_c$ in the QD ensemble are distributed according to Poisson statistics, so that $\langle N_g \rangle$ does not reach 0 when $n_c = 2$. Furthermore, entering the lasing regime requires consideration of photon losses in the cavity and Auger losses of the QDs. Here, we theoretically calculate the dependence of lasing threshold $\langle N_{las} \rangle$ on the Auger recombination (characterized by biexciton Auger lifetime $\tau_{A,XX}$) and the charging number per QD $\langle n_c \rangle$ (Fig. 1b), based on a modified gain-switching model that considers the Poisson distribution of $\langle n_c \rangle$ and the biexciton gain in charged QDs (Supplementary Note 1). Notably, when the QD is uncharged ($\langle n_c \rangle = 0$ in Fig. 1b), the threshold $\langle N_{las} \rangle$ changes less with increasing $\tau_{A,XX}$. This suggests that the suppressed Auger effect alone does not effectively lower the lasing threshold in the absence of QD charging. Similarly, when $\tau_{A,XX}$ is short, increasing $\langle n_c \rangle$ does not significantly lower the threshold. This is because QD charging exacerbates the Auger process, counteracting the threshold-lowering effect of multi-charged exciton gain. When $\tau_{A,XX}$ gradually prolongs, QD charging has a progressively more significant effect on threshold reduction. This implies that the Auger lifetime must be sufficiently long for the lowest $\langle N_{las} \rangle$ to be achieved by QD charging. Therefore, this theoretical analysis suggests that the relatively heavy charging and effective suppression of Auger recombination in QDs have to be satisfied simultaneously to achieve a lower lasing threshold.

**Design and characterizations of CsPbBr₃ QRs**

On the basis of the above theoretical analysis, we design one-dimensional perovskite

QRs to suppress the Auger process and to achieve a multi-charge charging via a state engineering strategy (Methods). Firstly, to significantly reduce Auger recombination, we elongate the QDs to construct QRs with a higher aspect ratio (Fig. 1c). In this case, the Auger recombination is reduced because the exciton motion becomes diffusive and their collisions are diminished[21,22], and, spatial-confined e-h pairs are transformed into Coulombically bound excitons[1,23]. It is worth noting that the QRs still have the same twofold-degenerated band edges as QDs (Supplementary Note 2), which means that QRs still retain the excellent properties of QDs as a laser gain medium. In addition, perovskite QRs are more advantageous than conventional CdSe-based QR because perovskite QRs are defect tolerant whereas CdSe QR are susceptible to large surface trapping[24,25]. Secondly, to achieve QR charging with sufficient electrons in a nondestructive manner, we design QRs with a self-charging effect by introducing bromine interstitial ($Br_i^-$) states (Fig. 1d) as hole acceptors to leave extra electrons in the conductive band (Supplementary Fig. 5b(left)). For this, a CsBr-rich environment is required during the QR self-assemble process, as the formation of $Br_i^-$ states are most energetically favorable under this condition[26]. Note that no self-charging effect presents in conventional-synthesized QRs without a CsBr-rich environment (Supplementary Note 3).

Detailed preparation of $CsPbBr_3$ perovskite QRs is presented in Methods, and their fundamental properties are presented in Supplementary Figs. 6a-d. The length and width of the QRs are 32 ± 4.8 nm and 14 ± 1.5 nm, respectively. An average absorption cross-section σ of $1.58 \times 10^{-13}$ cm$^2$ is at least an order of magnitude larger than that of previously reported QDs[27–29], which is beneficial to lasing. The PL characteristics of $CsPbBr_3$ QRs are revealed by single-dot spectroscopy, which provides quantitative access to the charging number, Auger lifetimes, biexciton quantum yields (QYs), and PL stability of QRs at the single-particle level. A typical PL intensity trajectory shows an increase in PL intensity over time (Fig. 1e), contrary to the commonly reported decrease in intensity[30]. The anomalously increased PL intensity is due to the QR transforming from charged (red region) to neutral (blue region) with increasing excitation time, which is confirmed by the radiative lifetime ratio between the two different PL regions (Supplementary Figs. 6f-i, Supplementary Note 6). By constructing a charging-associated Auger recombination model (Supplementary Note 7), we calculate the evolution of the charging number $n_c$ and nonradiative Auger lifetime of biexciton ($\tau_{A,XX^{n_c-}}$) with time (Fig. 1f). It is revealed that the QR possesses ~20 charges initially (blue stars), demonstrating an excellent self-charging effect. Nevertheless, under continuous illumination, these charges vanish due to the loss of hole acceptors. As $n_c$ decreases, $\tau_{A,XX^{n_c-}}$ prolongs due to the reduction of Auger pathways (red circles). Remarkably, the QR displays an Auger lifetime of 13.9 ns for neutral biexciton ($n_c = 0$, red solid circle), which is two orders of magnitude longer than that of conventional CdSe-based QDs and perovskite QDs (30 - 300 ps)[1,29]. The longer Auger lifetime implies a significant suppression of Auger recombination, and the significantly suppressed Auger recombination enables high PLQYs of charged single exciton (90% in Supplementary Fig. 6e) and biexciton (98% in Supplementary Fig. 6f), and sufficient $n_c$ in QRs. Both the self-charging effect and suppressed Auger process of QRs would contribute to the achievement of lasing with ultra-low threshold. Nevertheless, under continuous illumination, these charges gradually disappear due to the loss of hole acceptors. This is because the $Br_i^-$ states cannot exist stably in the QRs due to the self-reorganization tendency[26], where $Br^-$ naturally migrates to the QR surface and occupy their natural position ($Br_{Br}^-$) to eliminate the $Br_i^-$ states (Supplementary Fig. 5e). Therefore, the self-charged QRs cannot be stably charged with multiple charges in this case.

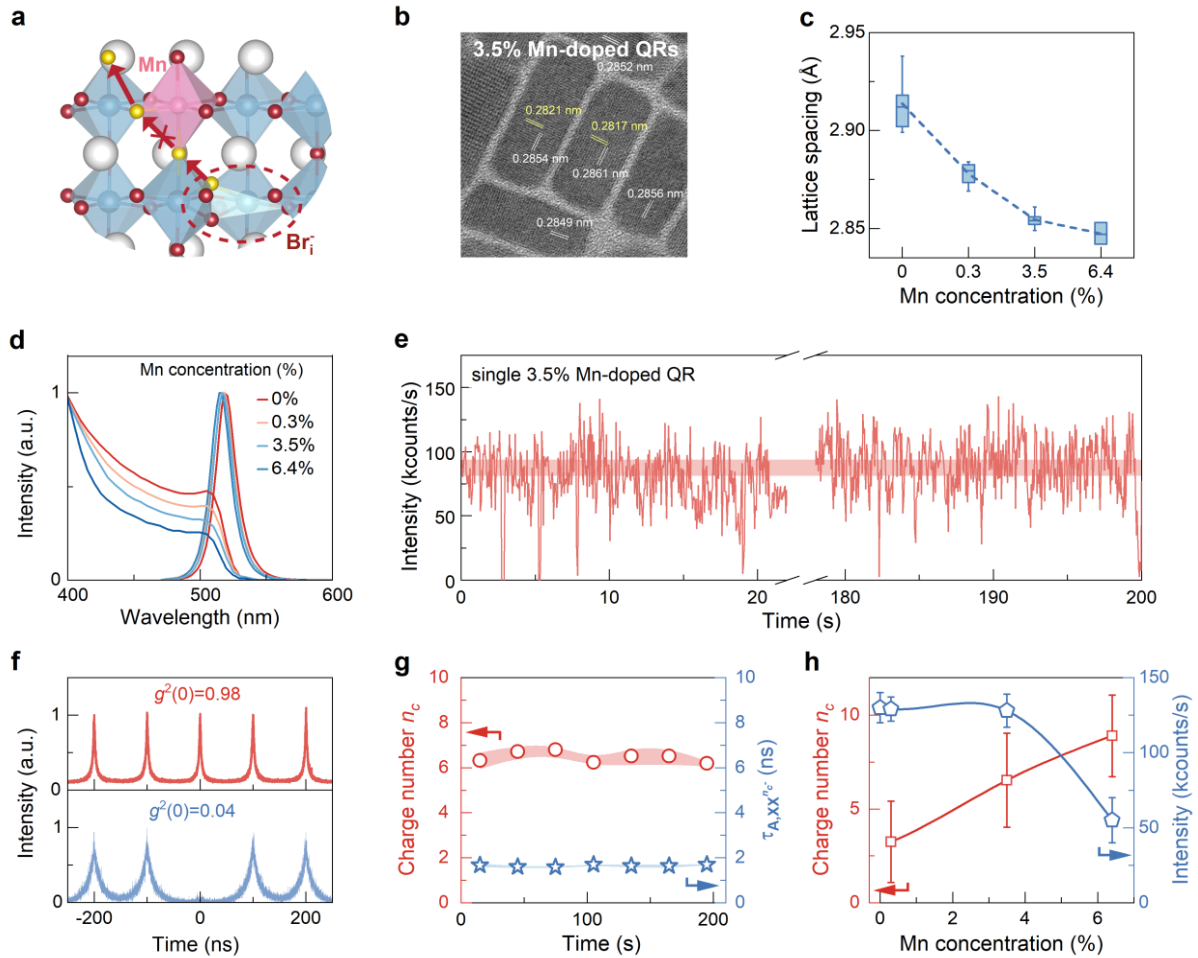

**Fig. 2 | Characteristics of Mn-doped CsPbBr₃ perovskite QRs.** (a) Schematic depiction of bromine interstitial ($Br_i^-$) state migration forbidden by Mn-doping-increased energy barrier, which maintains the presence of hole acceptors to stabilize QR multi-charge self-charging. (b) High-resolution transmission electron microscopy (HRTEM) images of Mn-doped CsPbBr₃ QRs with Mn concentration of 3.5%. (c) Lattice spacing of CsPbBr₃ QRs as a function of Mn doping concentration, showing lattice contraction with increasing Mn doping concentration.

(d) Absorption and PL spectra of QRs with different Mn doping concentrations dispersed in cyclohexane. (e) Typical PL intensity time trajectory for single 3.5% Mn-doped QRs at excitation condition $\langle N \rangle = 0.2$. (f) Corresponding original second-order correlation function $g^2$ curve (upper panel) and time-gated $g^2$ curve (lower panel) of the single QR. The $g^2(0)$ value of the time-gated $g^2$ curve is 0.1, which is less than 0.5, confirming that the observed PL is derived from a single QR. (g) Evolutions of charging number $n_c$ (red circles) in the QR and corresponding biexciton Auger lifetime $\tau_{A,XX^{n_c-}}$ (blue stars) over time. Shaded areas indicate errors. (h) Average PL intensity (blue pentagons) and charging number $n_c$ (red squares) of ~250 single QRs with different Mn-doping concentrations.

**Mn-doping enables stable multi-charge charging**

To address this issue of unstable charging, we develop a Mn-doping strategy to block $Br^-$ migration along the octahedrons of perovskite (Fig. 2a; Methods). This is because Mn possesses the largest number of empty orbitals, and these empty orbitals can invite the lone-pair electron of Pb-Br octahedrons during the hybridization of its 3d orbital with Pb 6s and Br 4p. The hybridization strengthens the chemical bond in surrounding Pb-Br octahedrons to increase the energy barrier for migration.[31] Therefore, Mn dopant is the best candidate. As a result, the $Br_i^-$ states as hole acceptors can stably exist in the Mn-doped QRs to achieve stable multi-charge charging.

Experimentally, we dope QRs with Mn using different concentrations (0.3%, 3.5% and 6.4%). The high-resolution TEM images (Fig. 2b, Supplementary Fig. 13a) reveal the lattice contraction in Mn-doped QRs (Fig. 2c), confirming the successful incorporation of Mn into the QR lattice, which is also supported by the X-ray diffraction patterns (Supplementary Fig. 13b). Absorption and PL spectra of QRs with different Mn doping concentrations are present in Fig. 2d. The gradual decrease of the first excitonic absorption peaks in absorption spectra indicates an increase in the charging numbers with increasing Mn-doping concentrations in the QRs.[32] The PL properties of single 3.5% Mn-doped CsPbBr$_3$ QRs are shown as an example (the others in Supplementary Fig. 14). The typical PL intensity trajectory remains stable over time (Fig. 2e), in sharp contrast to that of undoped QRs (Supplementary Fig. 7). The $g^2(0)$ value of the original second-order correlation function $g^2$ curve is still as high as 0.98 (Fig. 2f), which is significantly higher than that of the conventional single QD (~ 0.1)[33–36], is mainly contributed by the efficient biexciton emission[37] due to the effectively reduced Auger recombination in Mn-doped QRs. Based on the lifetimes of charged exciton ($X^{n_c-}$) and charged biexciton ($XX^{n_c-}$) obtained by fitting their decay curves (Supplementary Fig. 15), the charging number $n_c$ as well as Auger lifetime $\tau_{A,XX^{n_c-}}$ in the Mn-doped QR can be calculated, as shown in Fig. 2g (detailed in Supplementary Note 7). It is found that ~6 stable charges are obtained in the 3.5% Mn-doped QRs, and $\tau_{A,XX^{n_c-}}$ remains at 1.7 ns due to charging. Although Auger lifetime decrease from 13.9 ns to 1.7 ns by 6-charge charging, the lifetime of 1.7 ns is still comparable to that of neutral conventional perovskite QDs[28,29] or Auger-engineered II-VI QDs[3,38]. If conventional QDs are charged with 6 electrons, their Auger lifetimes will be shortened to tens of picoseconds (according to Supplementary Fig. 12). Therefore, a combination of Auger suppression and charging number must be considered to achieve a lower lasing threshold.

To determine the optimal Mn-doping concentration, we consider the effect of

Mn-doping concentrations on charging numbers $n_c$ and PL intensities. The average $n_c$ increases with increased Mn concentration (Fig. 2h), which is consistent with the results of absorption spectra of QRs in Fig. 2d. In addition, we do not observe significant change in PL intensity at Mn-doping concentrations of 0.3% and 3.5% as compared to the undoped QRs, while PL intensity drops significantly at the doping concentration of 6.4% (Fig. 2h), as excess Mn doping can introduce defects to reduce the PL intensity[39]. In addition, $n_c$ of more than 8 in the QR has a negligible effect on threshold reduction according to our theoretical analysis (Supplementary Fig. 4). As a result, the 3.5% Mn-doped QRs that exhibit high PL intensity and considerable high $n_c$ are thus determined as the optimal gain media.

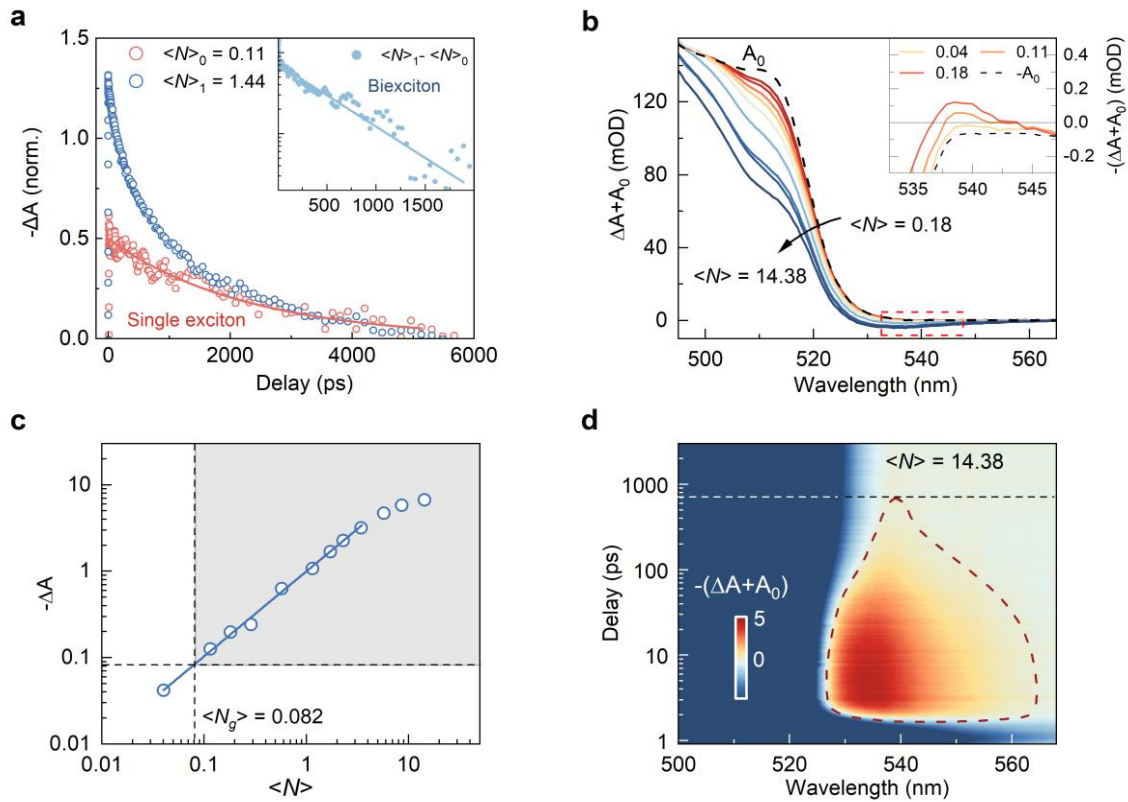

**Fig. 3 | Femtosecond transient absorption (TA) and optical gain of Mn-doped CsPbBr$_3$ QRs.** (a) By subtracting the exciton kinetic curve of $\langle N \rangle = 0.11$ (red circles) from that of

$\langle N \rangle = 1.44$ (blue circles), the biexciton kinetic curve (inset) is obtained. The kinetic curves of both single exciton and biexciton are fitted by single-exponential functions (solid lines). (b) Nonlinear absorption spectra ($\Delta A + A_0$) at 3 ps under various $\langle N \rangle$. Inset: amplified absorption spectra showing the transition from absorption to net gain. (c) Absorption bleaching ($-\Delta A$) as a function of $\langle N \rangle$ for QRs indicates the gain threshold $\langle N_g \rangle = 0.082$, where the $-\Delta A = A_0$. (d) A two-dimensional pseudo-color plot of the net gain spectra. The gain lifetime $\tau_g$ is 735 ps at 538 nm.

**Femtosecond transient absorption and optical gain of QRs**

To investigate the optical gain performance of QRs, we perform femtosecond transient absorption (TA) spectroscopy measurements at different $\langle N \rangle$ (see Methods). Different from single-dot spectroscopy, TA spectroscopy allows the contributions of Auger lifetime and QR charging to the optical gain to be studied at the ensemble level. Figure 3a shows the TA kinetics probed in the ground-state bleach at $\langle N \rangle = 0.11$ (red circles) and $\langle N \rangle = 1.44$ (blue circles), normalized to their long-lived tails. The biexciton kinetic curve (inset) is obtained by subtracting the TA kinetics of $\langle N \rangle = 0.11$ (single exciton only) from that of $\langle N \rangle = 1.44$ (single exciton and biexciton). By single-exponential fitting, the charged biexciton lifetime is obtained as 1.2 ns, reconfirming the well-reduced Auger process.

Figure 3b displays the excited-state absorption spectra ($\Delta A + A_0$), where $\Delta A$ is the TA spectra and $A_0$ is the steady-state absorbance spectrum. $\Delta A + A_0$ decreases progressively with increasing $\langle N \rangle$, and the optical gain occurs when $\Delta A + A_0 < 0$ in the low-energy region (gray region). The inset graph clearly shows the transition from optical absorption to optical gain.

To quantify the gain threshold $\langle N_g \rangle$, we also plot -ΔA as a function of $\langle N \rangle$ at 538 nm (Figure 3c). The linear dependence of -ΔA on $\langle N \rangle$, which deviates from the quadratic dependence of biexciton gain, is a strong indication of charged exciton gain[3,40]. The threshold $\langle N_g \rangle$ is determined to be 0.082 by extracting the horizontal coordinates of the fitted line when -ΔA=$A_0$. This quite low threshold is indicative of the multi-charge charging nature of QRs. In Figure 3d, we plot the time evolution of the net gain spectra at $\langle N \rangle = 14.38$, where the net gain area (-(ΔA+$A_0$) > 0) is represented by red dashed contour. The gain decays with delay time due to multi-exciton recombination, and the gain lifetime ($\tau_g$) is determined to be 735 ps at 538 nm. The competitive $\tau_g$, which benefits from reduced Auger recombination, allows the use of long-duration pulse as a pump source in QR lasing applications.

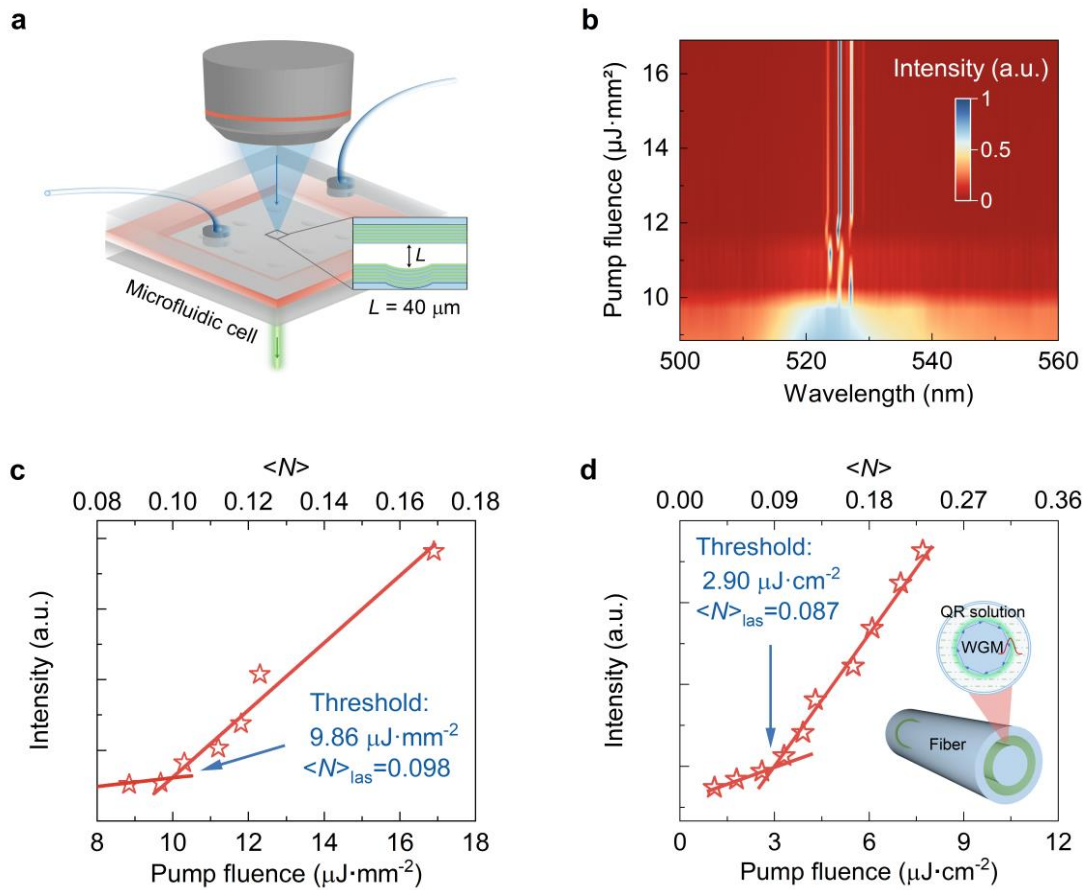

**Fig. 4 | Nanosecond and femtosecond pulse-pumped QR lasing.** (a) Schematic of the

Fabry–Pérot (F-P) microcavity structure. It is constructed by using one planar and one concave high-reflectivity (>99.9%) distributed Bragg reflector (DBR) mirrors facing each other. A microfluidic channel is formed between the two mirrors for the injection of QR solution. The cavity length $L$ is ~40 μm. (b) Normalization of the spectral peaks versus pump fluences highlighting the emergence of dominant laser modes with linewidths smaller than spontaneous emission. (c) Integrated intensity as a function of pump fluence, with the lasing threshold at 9.86 μJ/mm$^2$, corresponding to $\langle N \rangle = 0.098$. (d) Femtosecond pulse-pumped QR lasing: integrated intensity as a function of pump fluence, with the lasing threshold at 2.90 μJ/cm$^2$, corresponding to $\langle N \rangle = 0.087$. The inset shows the schematic of the whispering gallery mode (WGM) microcavity structure.

**Near-zero threshold liquid lasing**

QR liquid lasing using quasi-continuous nanosecond pulse pumping is shown in Figs. 4a-c. The Fabry-Pérot (F-P) microcavity is constructed using two distributed Bragg reflector (DBR) mirrors (Fig. 4a), and the transmission spectrum of the DBR mirrors is shown in Supplementary Fig. 16a, with the excitation and lasing wavelengths at 478 nm and 525 nm, respectively. Pump fluence-dependent emission spectra from the F-P microcavity filled with a low-concentration (0.09 μM) QR solution excited with 5 ns pump pulses (478 nm, 20 Hz), showing that a series of distinct narrow peaks emerge around 525.3 nm and grow rapidly with increasing pump fluences (Supplementary Fig. 16b). An obvious transition from spontaneous emission to lasing emission is demonstrated by the evolution of the normalized spectra with pump fluences, where the linewidth narrows significantly (Fig. 4b). A lasing linewidth at the pump fluence of 12.3 μJ/mm² is determined as 0.4 nm from a Lorentz fitting (Supplementary Fig. 16c). According to the central wavelength $\lambda$ of 525.3 nm and the lasing linewidth $\delta\lambda$ of 0.4 nm, the cavity quality factor $Q$ is derived to be ~1313 by using the formula $Q = \lambda/\delta\lambda$ [13,41]. The rapid increase in integrated intensity of the dominant emission peaks suggests the onset of the lasing regime (Fig. 4c). It leads to a well-defined threshold as low as 9.86 μJ/mm², which is lower than those of previously reported liquid lasers[6,7,11–13] (Supplementary Table S3).

For the multi-charged exciton gain mechanism, the lasing threshold is also expressed in terms of excitation condition $\langle N \rangle$ (Supplementary Fig. 16d). The threshold of 9.86 μJ/mm² corresponds to the $\langle N \rangle$ value of 0.098 for the 6-charged exciton gain, which means that 0.098 excitons per QR on average are sufficient for liquid lasing onset, significantly lower than the 1.15 excitons in conventional biexciton gain[3]. Notably, the threshold of 0.098 is lower than that of all previously reports on QD liquid lasers (Supplementary Table S3 for a detailed comparation) and all QD film-based lasers[1,4].

The record-low threshold is further confirmed by the femtosecond pulse-pumped QR lasing (Fig. 4d). Considering the inefficient optical confinement of the F-P microcavity, the capillary-tube whispering gallery mode (WGM) laser of QR solution is developed (inset). The WGM liquid lasing occurs with increasing pump fluence (Supplementary Fig. 17a), where a much higher Q-factor of ~3495 has been achieved. The lasing threshold is obtained as 2.9 μJ/cm² (Fig. 4d), which is again the lowest threshold for liquid lasers. The corresponding $\langle N \rangle$ of 0.087 at threshold (Supplementary Fig. 17c) reconfirm the near-zero threshold in QR lasing. To the best of our knowledge, this is the first time that a near-zero lasing threshold has been demonstrated.

## Conclusion

We develop stable self-charged perovskite quantum rods (QRs) via state engineering and Mn-doping strategy to reduce the lasing threshold by optimal combination of Auger recombination and charging number. A near-zero liquid lasing threshold of 0.098 is achieved using quasi-continuous nanosecond pulse pumping, which is the lowest threshold among all reported QD lasers. If the QR gain media is used for electrical-pumped laser, the threshold can be theoretically estimated to be as low as 3.79 A/cm$^2$ (Supplementary Note 8), which is ~700 times lower than that of a conventional QD laser (2700 A/cm$^2$). The results demonstrate the enormous potential of the specially engineered QRs in near-zero-threshold lasing applications.

## Supporting Information

Supplementary Figs. 1–17, Notes 1-9, Tables 1–4 and reference.

## Acknowledgements

The authors acknowledge financial support from the National Key Research and Development Program of China (No. 2022YFA1404201), the Natural Science Foundation of China (Nos. 62127817, U22A2091, U23A20380, 62325505, 62222509, 62274090, 62205187, 62305201, 62405211, W2412024 and 62375004), Program for Changjiang Scholars and

Innovative Research Team (No. IRT_17R70), Innovation Program for Quantum Science and Technology (Grant No. 2021ZD0302103), China Postdoctoral Science Foundation (No. 2022M722006), Fundamental Research Program of Shanxi Province (No. 202303021222031 and 202403021211012), Shanxi Province Science and Technology Innovation Talent Team (No. 202204051001014), Shanxi Province Science and Technology Major Special Project (No. 202201010101005), Shanxi "1331 Project", and 111 project (No. D18001).

**Methods:**

***Materials:*** Cesium carbonate ($Cs_2CO_3$, 99.0%), lead bromide ($PbBr_2$, 99.0%), tetraoctylammonium bromide (TOAB, 98.0%), manganese bromide ($MnBr_2$, 98.0%), 1-octadecene (ODE, >90%, GC), oleic acid (OA, AR), oleylamine (OAm, 80-90%), and cyclohexane (≥99.9%) were purchased from Aladdin. Polymethylmethacrylate (PMMA) and toluene (≥99.5%) were purchased from Sigma-Aldrich. Acetone were purchased from Sinopharm Chemical Reagent Co., Ltd. They were used as received without further purification.

***Preparation of Cesium-Oleate Solution:***

0.16 g $Cs_2CO_3$, 16 mL ODE and 1 mL OA were loaded in to a 50 mL three-neck flask. The mixture was dried at 120℃ for 30 min under vacuum, and then heated to 150℃ under $N_2$ to form a clear solution.

***Synthesis and Purification of $Cs_4PbBr_6$ NCs:***

The $Cs_4PbBr_6$ NCs were prepared as follows. 0.2 mmol $PbBr_2$, 10 mL ODE, 1 mL OA and 1 mL OAm were loaded into a 25 mL three-neck flask and dried at 120℃ for 30 min under vacuum. The mixture was then heated to 140℃ under $N_2$ to completely dissolve $PbBr_2$. 4.4 mL Cesium-oleate Solution (150℃) was then injected rapidly into the $PbBr_2$ solution. Ten seconds later, the flask was immersed into an ice water bath to terminated the reaction. By centrifugation at 8000 rpm for 5 min, solvents and excess reactants are discarded as supernatant. Then the precipitates were redissolved into 10 mL cyclohexane to form a $Cs_4PbBr_6$ NCs suspension. After another centrifugation at 3000 rpm for 5 min, a high-quality product was obtained for further use.

***Synthesis and Purification of $CsPbBr_3$ QRs:***

The synthesis of the $CsPbBr_3$ QRs was based on a water-triggered transformation process under ambient conditions. For conventional-synthesized QRs, 2 mL water and 2 mL 6 mM $Cs_4PbBr_6$ NCs solution were directly injected into a 10 mL vial. The mixture was left for 24 h to complete the transformation. For CsBr-rich QRs, we hope that $Br^-$ can enter the QRs to occupy the interstitial positions. For this purpose, 2 mL of the $Cs_4PbBr_6$ NCs solution was slightly injected into 2 mL CsBr solution (4 mM in water). For the purification of

CsPbBr$_3$ QRs, the CsPbBr$_3$ QRs obtained using two methods were centrifugated at 13000 rpm for 5 min, and the precipitates were redissolved in 2 mL cyclohexane to form CsPbBr$_3$ QRs suspensions. After another centrifugation at 8000 rpm for 5 min, high-quality CsPbBr$_3$ QRs were obtained for further use.

## *Doping Mn$^{2+}$ in CsPbBr$_3$ QRs:*

The postsynthesis Mn$^{2+}$ doping was processed under room temperature and ambient atmosphere. Typically, the Mn precursor solution was obtained by dissolving MnBr$_2$ in a mixture of acetone and toluene with volume ratio of 1: 3 by sonication. Then the prepared Mn$^{2+}$ precursor solution was added to the QRs solution (1 mL) under continuous stirring for 1 min. Different dopant concentrations were regulated by the amount of the precursor added into the QRs (Table S1). After doping, the obtained Mn-doped QRs were isolated by centrifuging at 13000 rpm for 5 min. After centrifugation, the supernatant was discarded and the precipitate was redispersed in 1 mL cyclohexane to obtained high-quality Mn-doped CsPbBr$_3$ QRs.

## *Sample Preparation for Single QR Spectroscopic Measurements:*

For single QR measurements, the cyclohexane solution of QRs was diluted and then spin-coated onto clean glass coverslips with a rotation speed of 3000 rpm for 1 min. PMMA film covering single QRs is prepared by spin coating to isolate and protect the particles.

## *F-P microcavity Fabrication:*

For lasing measurement, the F-P microcavity was constructed with one planar and one concave DBR mirror (dielectric coating with a maximum reflectivity ≥99.9% at a center wavelength of 525 nm) facing each other with a distance of 30 μm spaced by a copper. The fabrication of concave micro-structure was prepared by $CO_2$ laser.[42] A microfluidic channel was formed simultaneously, as shown in Figure 4a, where the solution of gain medium injected by a syringe pump can flow through the F-P microcavity structure.

For PL saturation curve measurements in liquid, mirrors of the F-P microcavity were replaced with uncoated transparent fused silica substrates. Other experimental parameters are the same as above.

*Characteristics:*

The size and lattice spacing of QRs were obtained from the TEM and high-resolution TEM image, respectively, using a JEM-2100 microscope. The absorption and PL emission spectra of the QRs in cyclohexane were recorded in a range of 300-800 nm by using a PerkinElmer Lambda 950 UV-VIS-NIR spectrometer and a Cary Eclipse Fluorescence Spectrophotometer, respectively. The mole concentration of $CsPbBr_3$ QRs and Mn dopants are determined by the inductively coupled plasma-mass spectrometry (ICP-MS, NexION350X, PerkinElmer, USA). XRD patterns were measured using a D2 PHASER diffractometer.

*Single-Dot PL Spectroscopic Measurements:*

A home-made confocal fluorescence imaging microscope was applied to collect the PL photons of single QRs. The single QRs were excited by a 439 nm pulsed laser (WL-SC-400-15-PP, NKT Photonics) with a repetition rate of 10 MHz. An oil-immersed high numerical aperture objective (Olympus, 100×, 1.3 NA) was used not only to focus a laser beam onto the QRs sample but also to collect PL simultaneously. The laser beam was sent into an inversion microscope (Olympus IX71) and reflected by a dichroic mirror (Semrock, Di03-R405), which finally focused on the sample by the objective. The PL, which was collected by the same objective, passed through the dichroic mirror and a high-pass filter (Semrock, BLP01-405R), and finally focused on a 100 μm pinhole to reject photons which are out of focus. Then the PL was splited by a 50/50 beam-splitter cube into two beams and detected by a pair of single-photon avalanche diode detectors (SPCM-AQR-15, PerkinElmer). PL intensity and lifetime information was recorded by a TTTR-TCSPC data acquisition card (HydraHarp 400, PicoQuant) with a temporal resolution of 16 ps and analyzed by a routine written in MATLAB. The scanning of the sample over the focused excitation spot by using a piezo-scan stage ((Piezosystemjena, T-405-01) with an active x-y-z feedback loop mounted on the inversion microscope. All measurements were performed at room temperature.

*Transient absorption (TA) measurements:*

TA measurements are performed on solutions of QRs in an airtight cuvette. A femtosecond fiber laser (YF-FL-10-IR/GN/UV, Hangzhou Yacto Technology) is used to generate 290 fs photon pulses with a wavelength of 1030 nm and a frequency of 20 kHz. The laser output is divided into two parts. The 343 nm pump pulse is generated from the main part by the triple-frequency methods. The reflected part is focused on a YAG crystal to generate a supercontinuum as a probe beam. The white light can be focused on the sample by an off-axis parabolic mirror. The probe beam passing through the sample is collected by a series of lenses and focused on a spectrometer. The time delay between the pump and probe beams is controlled by a motorized stage. The pump pulses are modulated by an asynchronous chopper and the transient signals are obtained by calculating the absorption changes of two adjacent probe pulses. All experimental data are corrected for the chirp induced by the nonlinear white light generation process and are performed at room temperature.

*Lasing measurements:*

For lasing measurements, the QR liquid lasers were pumped by vertically incident of a nanosecond pulsed laser (478 nm wavelength, 5 ns pulse width, 20 Hz repetition rate) or a femtosecond pulsed laser (400 nm wavelength, 100 fs pulse width, 1 kHz repetition rate). The pump energy can be varied using a neutral density attenuator. QRs emission signal in the cavity is collected with an optical fiber and then sent to a spectrometer (Horiba 320).